\documentclass[12pt]{article}
\usepackage[dvips]{graphicx}

\newcommand{\EM}	{{\rm em}}

\newcommand{\rad}	{{\rm rad}}
\newcommand{\external}	{{\rm ext}}
\newcommand{\internal}	{{\rm int}}
\newcommand{\trans}	{{\rm trans}}
\newcommand{\sing}	{{\rm sing}}
\newcommand{\Ptan}	{{P_{\tan}}}
\newcommand{\diag}	{{\rm diag}}
\newcommand{\BG}	{{\rm BG}}
\newcommand{\vac}	{{\rm vac}}

\newcommand{\BH}	{{\rm BH}}
\newcommand{\pl}	{{\rm pl}}
\newcommand{\maxi}	{{\rm max}}
\newcommand{\mini}	{{\rm min}}
\newcommand{\peak}	{{\rm peak}}
\newcommand{\rhopeak}	{{\rho\mbox{-}{\rm peak}}}
\newcommand{\Gpeak}	{{G\mbox{-}{\rm peak}}}

\newcommand{\lnear}	{{\begin{array}{c} < \\[-1.1em] \sim \end{array}}}
\newcommand{\gnear}	{{\begin{array}{c} > \\[-1.1em] \sim \end{array}}}

\newcommand{\fig}[1]	{Figure \ref{#1}}

\newcommand{\EQ}[1]	{(\ref{#1})}

\begin{document}

%%%% Preprint Numbers & Date %%%%
\begin{flushright}
 \begin{minipage}[b]{43mm}
  hep-th/0311133\\
  WIS/31/03-NOV-DPP\\
 \end{minipage}
\end{flushright}

%%%% Title & Author %%%%
\renewcommand{\thefootnote}{\fnsymbol{footnote}}
\begin{center}
 {\Large\bf
 Radiation Ball for a Charged Black Hole
 }\\
 \vspace*{3em}
 {Yukinori Nagatani}\footnote
 {e-mail: yukinori.nagatani@weizmann.ac.il}\\[1.5em]
 {\it Department of Particle Physics,\\
 The Weizmann Institute of Science, Rehovot 76100, Israel}
\end{center}
\vspace*{1em}

%%%% ABSTRACT %%%%
\begin{abstract}
 A radiation-ball solution 
 which is identified as a Reissner-Nordstr\"om black hole
 is found out.
 The radiation-ball,
 which is derived by analyzing the backreaction
 of the Hawking radiation into space-time,
 consists of radiation trapped in a ball
 by a deep gravitational potential
 and of a singularity.
 The Hawking radiation is regarded as
 a leak-out of the radiation from the ball.
 The gravitational potential becomes deep
 as the charge becomes large,
 however,
 the basic structure of the ball is independent of the charge.
 The extremal-charged black hole corresponds with
 the fully frozen ball by the infinite red-shift.
 The total entropy of the radiation in the ball,
 which is independent of the charge,
 obeys the area-law
 and is near the Bekenstein entropy.
\end{abstract}

% 04.70.Dy Quantum aspects of black holes, evaporation, thermodynamics
%
%\pacs{04.70.Dy}

%%%% Introduction %%%%
\newpage
\section{Introduction}\label{intro.sec}

Quantum mechanical properties of a black hole ---
the Hawking radiation \cite{Hawking:1975sw,Hawking:1974rv}
and the Bekenstein entropy \cite{Bekenstein:1973ur} ---
are most interesting and mysterious in modern field theory.
One of natural approaches
is considering the structure of the black hole.
The membrane-descriptions \cite{Thorne:1986iy},
the stretched horizon models \cite{Susskind:1993if,Susskind:1994sm},
the Planck solid ball model \cite{{Hotta:1997yj}}
and the quasi-particles model \cite{Iizuka:2003ad}
were proposed as this kind of approach.

Recently the radiation-ball structure
which can be identified as the Schwarzschild black hole
with the quantum properties
was derived by investigating the backreaction of the Hawking radiation
into space-time \cite{Nagatani:2003rj}.
The structure consists of radiation and of a singularity.
The radiation is trapped into the ball
by a deep gravitational potential
and has the Planck temperature.
The gravitational potential is self-consistently produced.
The exterior of the ball is fully corresponding
with the Schwarzschild black hole.
The Hawking radiation is reproduced as
a leak-out of the radiation from the surface of the ball.
The information paradox of the black hole does not arise
because the structure has no horizon.
The Bekenstein entropy is regarded as being
carried by the internal radiation of the ball
because
the total entropy of the radiation
is proportional to the surface-area of the ball
and is near the Bekenstein entropy.

The D-brane descriptions of the (near) extremal-charged black hole
succeeded in deriving the entropy
with the correct coefficient of the area-law \cite{Strominger:1996sh}
and the Hawking radiation \cite{Horowitz:1996fn}.
These approaches are just considering the dual structure of the black hole.
To consider the relation between the radiation-ball
and the D-brane description,
we should derive the radiation-ball for the charged black hole.

In this paper we will extend the radiation-ball
to a charged black hole.
We assume that the only singularity carries the charge
and the radiation is neutral.
We will find that
the basic structure of the radiation-ball is not changed
by the extension.
The exterior of the ball corresponds
with the Reissner-Nordstr\"om black hole.
The area-law of the entropy is also not changed.
The effect of the extension mainly arises
in the gravitational potential in the ball.
The gravitational potential in the ball becomes deep
as the charge becomes large.
In the case of the extremal-charged black hole,
the radiation-ball is fully frozen up
for a observer at the infinite distance
because the red-shift effect becomes infinitely large.

The plan of the rest of the paper is the following.
In the next section % \ref{backreaction.sec}
we present a framework of the investigation.
In Section \ref{solution.sec},
we derive the solution of the radiation-ball.
In Section \ref{property.sec},
we consider the properties of the ball.
In the final section we give discussions.

\section{Framework}\label{backreaction.sec}

We will consider the spherically symmetric static space-time
parameterized by the time coordinate $t$
and the polar coordinates $r$, $\theta$ and $\varphi$.
The generic metric is 
\begin{eqnarray}
 ds^2
  &=&
  F(r) dt^2 - G(r) dr^2 - r^2 d\theta^2 - r^2 \sin^2\theta \, d\varphi^2,
  \label{generic-metric}
\end{eqnarray}
where
the elements $F(r)$ and $G(r)$ are functions depending only on $r$.
In the space-time
the solution of the spherically symmetric static electromagnetic field 
becomes
${\cal F}
= E(r) \sqrt{FG} dt \wedge dr
+ B(r) r^2\sin\theta d\theta \wedge d\varphi$,
where
\begin{eqnarray}
 E(r) &:=& \frac{Q_{\rm e}}{r^2}, \qquad B(r) \ :=\  \frac{Q_{\rm m}}{r^2}
  \label{EM-field}
\end{eqnarray}
are the radial elements of the field.
$Q_{\rm e}$ and $Q_{\rm m}$
are the electric charge and the magnetic charge of the center ($r=0$)
respectively.
We define $Q^2 := Q_{\rm e}^2 + Q_{\rm m}^2$.

When we do not consider the backreaction of the Hawking radiation
and assume the asymptotic flatness,
the space-time becomes the Reissner-Nordstr\"om (RN) black hole.
The elements of the RN metric are given by:
\begin{eqnarray}
 F_\BH(r) &=&
  \left(1 - \frac{r_\BH}{r}\right)
  \left(1 - q^2 \frac{r_\BH}{r}\right),
  \label{RN-F}\\
 G_\BH(r) &=& 1/F_\BH(r),
  \label{RN-G}
\end{eqnarray}
where $r_\BH$ is the radius of the outer horizon and
$q := Q/(m_\pl r_\BH)$ is a charge-parameter
which has a value from 0 to 1.

% then both the density \EQ{rho} and the pressures (\EQ{Pr} and \EQ{Ptan})
% diverge on the horizon $r = r_\BH$.
% %
% The total radiation-energy around the black hole
% $ \int_{r_1}^{r_2} 4 \pi r^2 \rho(r) = 1/(r_1 - r_\BH) + \cdots$
% also diverges as $r_1 \rightarrow r_\BH$.
% %
% This problem requires that we should consider
% the backreaction of the radiation into the space-time structure.

The Hawking temperature of the charged black hole becomes
\begin{eqnarray}
  T_\BH &=& \frac{1}{4\pi} \frac{1}{r_\BH} (1 - q^2).
\end{eqnarray}
When $q=1$, the black hole becomes the extremal
and its Hawking temperature becomes zero.
We put the black hole into the background of the radiation
with the temperature $T_\BH$
to consider the stationary situation and the equilibrium of the system
\cite{Gibbons:1976ue}.
The energy density of the background radiation ($r\rightarrow\infty$)
is given by the thermodynamical relation:
\begin{eqnarray}
 \rho_\BG &=& \frac{\pi^2}{30} g_* T_\BH^4,
  \label{rho-BG}
\end{eqnarray}
where $g_*$ is the degree of the freedom of the radiation.
We assume that $g_*$ is a constant for simplicity.

Here we should introduce the positive cosmological constant
$\Lambda = (8\pi/m_\pl^2) \rho_\vac$
to stabilize the background universe from the effect of 
constant energy density $\rho_\BG$.
The background space-time becomes the Einstein static universe 
by choosing $\rho_\vac = \rho_\BG$.

In the situation
we expect the local temperature distribution of the radiation as
\begin{eqnarray}
 T(r) &:=& \frac{T_\BH}{\sqrt{F(r)}} \label{Local-Temperature}
\end{eqnarray}
due to the effect of the gravitational potential $F(r) = g_{tt}(r)$
\cite{Hotta:1997yj,Iizuka:2003ad,Nagatani:2003ps}.
Therefore the energy-density and the pressures of the radiation become
\begin{eqnarray}
 \rho(r)  &=&             \frac{\rho_\BG}{F^2(r)} \label{rho} \\
 P_r(r)   &=& \frac{1}{3} \frac{\rho_\BG}{F^2(r)}, \label{Pr} \\
 \Ptan(r) &=& \frac{1}{3} \frac{\rho_\BG}{F^2(r)}, \label{Ptan}
\end{eqnarray}
respectively.
$P_r(r)$ is the pressure in the $r$-direction and
$\Ptan(r)$ is the pressure in the tangential direction
($\theta$- and $\varphi$- direction).

\section{Radiation-Ball Solution}\label{solution.sec}

The space-time structure of the black hole
including the backreaction from the radiation
is computed by solving the Einstein equation
\begin{eqnarray}
 R_\mu^{\ \nu} -\frac{1}{2} R \delta_\mu^{\ \nu} - \Lambda\delta_\mu^{\ \nu}
 &=&
 \frac{8\pi}{m_\pl^2}
 \left\{ T_{\rad}{}_{\mu}^{\ \nu} + T_{\EM}{}_{\mu}^{\ \nu} \right\}
\end{eqnarray}
for the metric \EQ{generic-metric}.
Elements of the energy-momentum-tensor for the radiation
$T_{\rad}{}_{\mu}^{\ \nu}(r)
= \diag\left(\; \rho(r), -P_r(r), -\Ptan (r), -\Ptan(r) \; \right)$
are given by \EQ{rho}, \EQ{Pr} and \EQ{Ptan}.
$T_{\EM}{}_{\mu}^{\ \nu}$ is the energy-momentum-tensor for the
electromagnetic field \EQ{EM-field}.
%
% in \EQ{Erho}, \EQ{EPr} and \EQ{EPtan}
%
The Einstein equation becomes following three equations:
\begin{eqnarray}
  \frac{-G + G^2 + r G'}{r^2 G^2}
   &=&
   \frac{8 \pi}{m_\pl^2}
   \left\{\rho(r) + \rho_\BG + \frac{Q^2}{8\pi} \frac{1}{r^4} \right\},
   \label{Erho}\\
  \frac{F - F G + r F'}{r^2 F G}
   &=&
   \frac{8 \pi}{m_\pl^2}
   \left\{P_r(r) - \rho_\BG - \frac{Q^2}{8\pi} \frac{1}{r^4} \right\},
   \label{EPr}\\
  \frac{
   - r (F')^2 G  - 2 F^2 G'
   - r F F' G'
   + 2 F G (F' + r F'')
  }{ 4 r F^2 G^2}
   &=&
   \frac{8 \pi}{m_\pl^2}
   \left\{\Ptan(r) - \rho_\BG + \frac{Q^2}{8\pi} \frac{1}{r^4} \right\},
   \label{EPtan}
\end{eqnarray}
where $m_\pl$ is the Planck mass.
We obtain the relation
\begin{eqnarray}
 G(r) &=& \frac{1 - \frac{r}{2} \frac{\rho'(r)}{\rho(r)}}
  {1 + \frac{8\pi}{m_\pl^2} r^2
   \left\{
    \frac{1}{3} \rho(r) - \rho_\BG - \frac{Q^2}{8\pi}\frac{1}{r^2}
   \right\} }
  \label{G-rho}
\end{eqnarray}
from the equation \EQ{EPr} with \EQ{Pr}.
By substituting \EQ{G-rho} into the equation \EQ{Erho} with \EQ{rho},
we obtain the differential equation for the energy density $\rho(r)$ as
\begin{eqnarray}
 	&& 
		\:-\: 24 r 		\rho^2
		 \left\{ \textstyle
		  \rho  - \rho_\BG  + \frac{Q^2}{8\pi}\frac{1}{r^4}
		 \right\}
		\:+\: 12	r^2	\rho\rho'
		 \left\{ \textstyle
		  \rho + 2 \frac{Q^2}{8\pi}\frac{1}{r^4}
		 \right\}
	\nonumber\\
	&&
		\:+\:	r^3	\rho'^2
		 \left\{ \textstyle
		  \rho - 9 \rho_\BG - 9 \frac{Q^2}{8\pi}\frac{1}{r^4}
		 \right\}
		\:-\: 2 	r^3	\rho	\rho''
		 \left\{ \textstyle
		  \rho -3 \rho_\BG -3 \frac{Q^2}{8\pi}\frac{1}{r^4}
		 \right\}
	\nonumber\\
	&& \:+\: \frac{3 m_\pl^2}{8\pi}
	\left\{
		- 4 \rho \rho'
		+ 3 r \rho'^2
		- 2 r \rho \rho''
	\right\} \;=\; 0. \label{rhoEQ}
\end{eqnarray}
We also obtain the same differential equation \EQ{rhoEQ}
by substituting \EQ{G-rho} into the equation \EQ{EPtan} with \EQ{Ptan},
therefore,
the Einstein equation in \EQ{Erho}, \EQ{EPr} and \EQ{EPtan}
and the assumption of the energy-momentum tensor in
\EQ{rho}, \EQ{Pr} and \EQ{Ptan} are consistent.
%
% Although the number of the differential equations exceeds
% the number of the unknown functions, there exists a solution.
% %
% This is a non-trivial feature of the system.

\begin{figure}
 \begin{center}
  \includegraphics[scale=0.8]{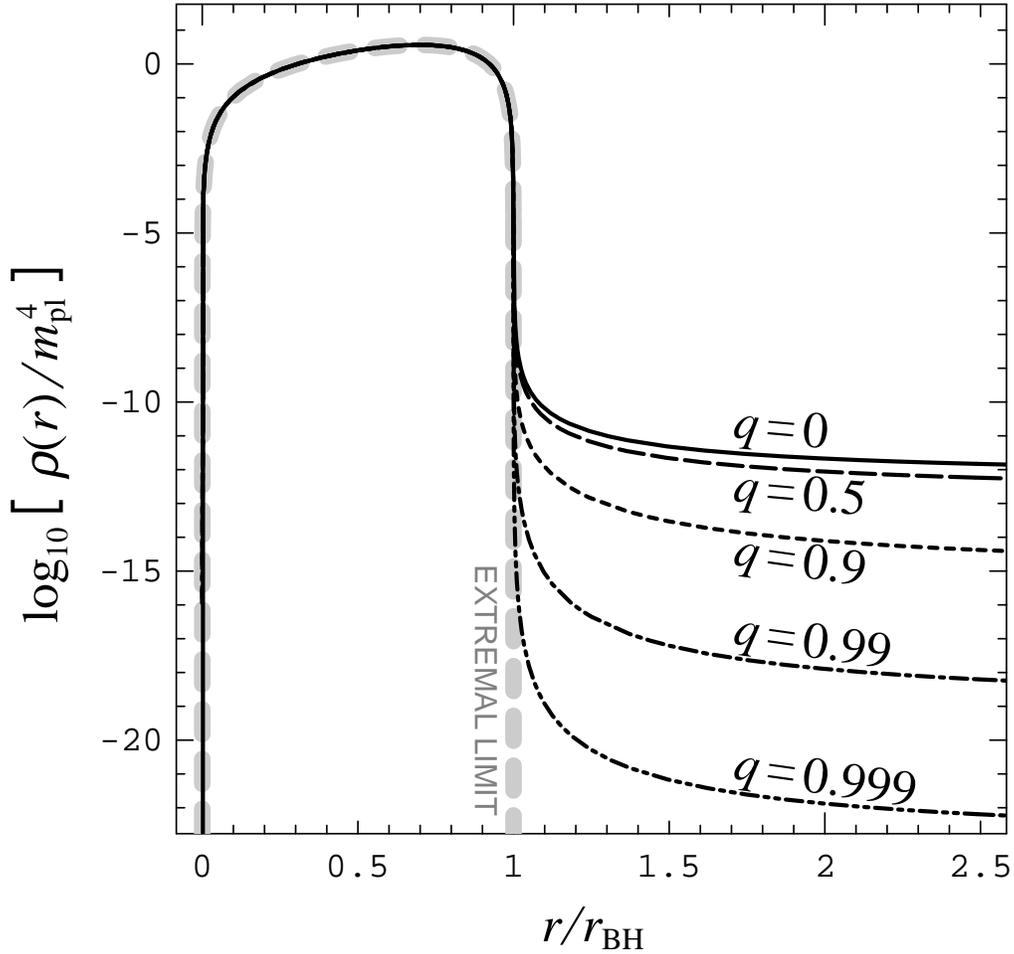}%
 \end{center}
 \caption{%
 Distributions of the radiation-energy-density
 $\rho(r)$ in the solution of the radiation-ball
 for the radius $r_\BH = 100\times l_\pl$, $g_* = 4$
 and for various charges $q$.
 The horizontal axis is the coordinate $r$
 normalized by $r_\BH$.
 The solutions for $q := Q/(r_\BH m_\pl) = 0, 0.5, 0.9, 0.99$ and
 $0.999$ are displayed.
 The thick dotted gray curve is
 the solution for the extremal limit ($q \rightarrow 1$).
 We find that the distribution in the ball ($r < r_\BH$)
 has a universal form.
 The density $\rho(r)$ approaches to the background density $\rho_\BG$
 defined in \EQ{rho-BG} when $r$ becomes large.
 \label{RHOLogMult.eps}%
 }%
\end{figure}

\begin{figure}
 \begin{center}
  \includegraphics[scale=0.8]{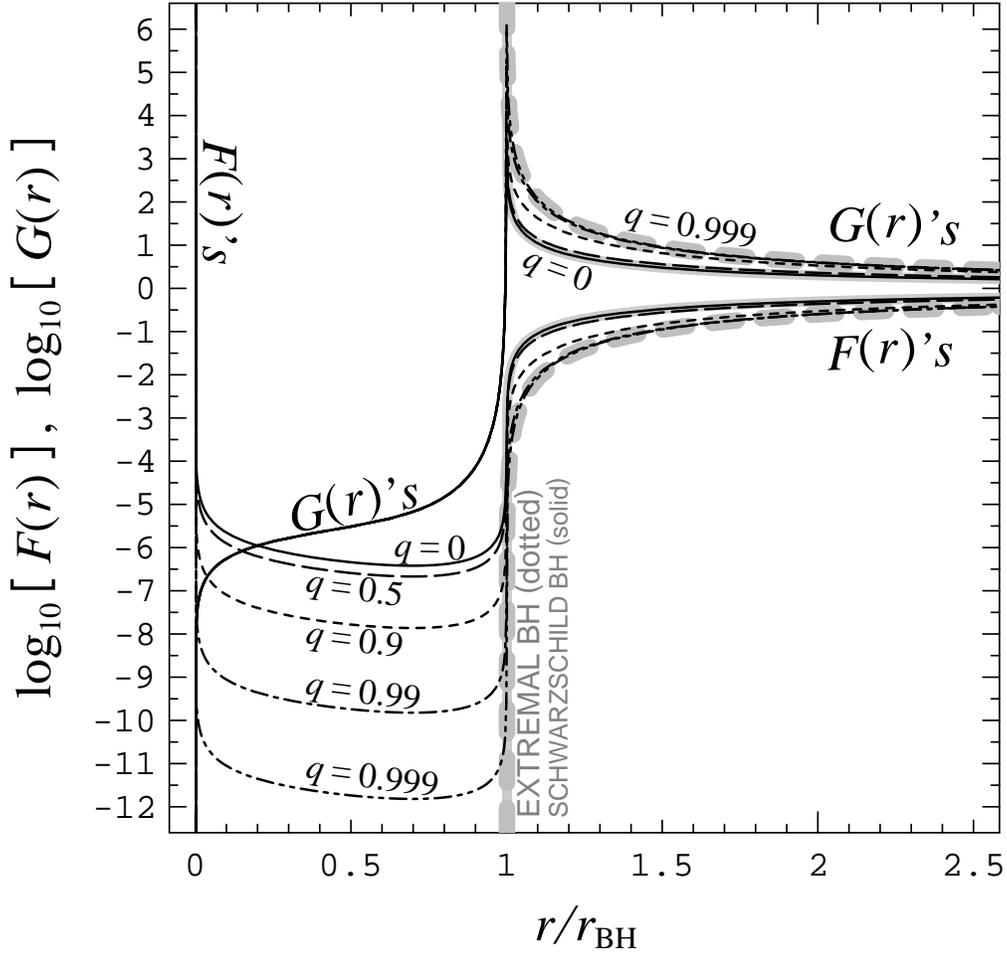}%
 \end{center}
 \caption{%
 Elements of the metric $F(r)$ and $G(r)$ in the solution
 of the radiation-ball
 with $r_\BH = 100 \times l_\pl$, $g_* = 4$ and
 for various charges $q = 0, 0.5, 0.9, 0.99$ and $0.999$.
 The thick dotted gray curves indicate the metric elements $F_\BH(r)$
 and $G_\BH(r)$ of the exterior part of the extremal-charged RN black hole
 ($q = 1$).
 The thick solid gray curves indicate the exterior part of
 the Schwarzschild metric ($q = 0$).
 We find that the function form of $G(r)$ in the ball
 is almost independent of $q$.
 \label{FG.eps}%
 }%
\end{figure}

The differential equation \EQ{rhoEQ} is numerically solved
and we find out the solution
whose exterior part corresponds with the charged black hole
in the Einstein static universe.
The Mathematica code for the numerical calculation
can be downloaded on \cite{MathCode:2003}.
The solution is parameterized by $r_\BH$ and $q$
which are the outer horizon radius and the charge-parameter
of the correspondent RN black hole respectively.
The numerical solutions for various $r_\BH$ are displayed in
\fig{RHOLogMult.eps}.
The element of the metric $F(r)$ is derived by \EQ{rho}
and $G(r)$ is derived by \EQ{G-rho}.
Typical forms of the metric elements are displayed in \fig{FG.eps}.
The solution indicates that
most of the radiation is trapped in the ball
by the gravitational potential $F(r)$ and
the radius of the ball is given by $r_\BH$.
This is certainly the charge-extension of the radiation-ball
\cite{Nagatani:2003rj}.

\section{Properties of the Radiation-Ball}\label{property.sec}

\subsection{Exterior of the Radiation-Ball}

For the external region ($r \gnear r_\BH + l_\pl$)
the differential equation \EQ{rhoEQ}
is approximated to
\begin{eqnarray}
	&&
	\frac{Q^2}{8\pi}\frac{1}{r^4}
	\left\{
	-24 r   \rho^2
	+24 r^2 \rho \rho'
	- 9 r^3 \rho'^2
	+ 6 r^3 \rho \rho''
	\right\} \nonumber\\
	&& +
	\frac{3 m_\pl^2}{8\pi}
	\left\{
	- 4 		\rho		\rho'
	+ 3 	r	\rho'^2
	- 2 	r	\rho		\rho''
	\right\}
	\ =\  0
\end{eqnarray}
when $r_\BH$ is much greater than the Planck length $l_\pl := m_\pl^{-1}$.
The approximated equation has a solution
\begin{eqnarray}
 \rho_\external(r)
  &=&
  \rho_\BG  \times
  \left(1 - \frac{r_\BH}{r}\right)^{-2}
  \left(1 - q^2\frac{r_\BH}{r}\right)^{-2},
\end{eqnarray}
and the elements of the metric become
\begin{eqnarray}
 F_\external(r)
  &=& \left(1 -    \frac{r_\BH}{r}\right)
      \left(1 - q^2\frac{r_\BH}{r}\right),
  \label{Fexternal}\\
 G_\external(r)
  &=&
  \left[
   F_\external(r)
      + \frac{8\pi}{3 m_\pl^2} r^2 \rho_\BG
	\frac{F_\external^{-1}(r) - 3 F_\external(r)}
	{1 - q^2 \frac{r_\BH^2}{r^2}}
  \right]^{-1}. \label{Gexternal}
\end{eqnarray}
The approximated solution $\rho_\external(r)$
is corresponding to \EQ{rho} with the RN metric \EQ{RN-F}.
The resultant metric ($F_\external(r)$ and $G_\external(r)$) is also
consistent with the external part of the RN metric (\EQ{RN-F} and \EQ{RN-G})
with a background correction of the Einstein static universe.

\subsection{Interior of the Radiation-Ball}

For the internal region ($l_\pl \lnear r \lnear r_\BH - l_\pl$)
the differential equation \EQ{rhoEQ}
is approximated to
\begin{eqnarray}
 	-24		\rho(r)^2
	\;+\;12	r	\rho(r)	\rho'(r)
	\;+\;	r^2	\rho'(r)^2
	\;-\;2 	r^2	\rho(r)	\rho''(r)
	&=& 0 \label{RhoOutEQ}
\end{eqnarray}
for $r_\BH \gg l_\pl$.
We obtain the solution of the equation \EQ{RhoOutEQ} as
\begin{eqnarray}
 \rho_\internal(r)
  &=&
  \frac{135}{\pi} \frac{1}{g_*} m_\pl^4
  \left(\frac{r}{r_\BH}\right)^2
  \left[ 1 - \left(\frac{r}{r_\BH}\right)^5 \right]^2,
\end{eqnarray}
where the coefficients of the solution are determined by
matching to the numerical solution.
This does not depend on the charge $q$
because $T_\EM{}_\mu^{\ \nu} \sim q^2 m_\pl^2/r_\BH^2$
is much smaller than $T_\rad{}_\mu^{\ \nu} \sim m_\pl^4$ in the region.
The elements of the metric become
\begin{eqnarray}
 F_\internal(r)
  &=&
  \frac{g_*}{720 \sqrt{2\pi}}
  \frac{(1 - q^2)^2}{m_\pl^2 r_\BH^2}
  \left( \frac{r_\BH}{r} \right)
  \left[1 - \left(\frac{r}{r_\BH}\right)^5\right]^{-1},\\
 G_\internal(r)
%  &=&
%   \left[ \left(\frac{r_\BH}{r}\right)^5 - 1 \right]^{-1}
%   \left[
%    \frac{1}{5}
%    + \frac{72}{g_*} m_\pl^2 r_\BH^2 \left(\frac{r}{r_\BH}\right)^4
%      \left\{1 - \left(\frac{r}{r_\BH}\right)^5 \right\}^2
%   \right]^{-1} \nonumber\\
  &\simeq&
   \frac{g_*}{72}
   \frac{1}{m_\pl^2 r_\BH^2}
   \left(\frac{r}{r_\BH}\right)
   \left[1 - \left(\frac{r}{r_\BH}\right)^5 \right]^{-3}.
\end{eqnarray}
$G_\internal(r)$ does not depend on $q$.
$F_\internal(r)$ is proportional to $(1-q^2)^2$.
When we consider the extremal limit $q \rightarrow 1$,
$F_\internal(r)$ goes to zero, namely,
the red-shift of the radiation-ball goes to infinity
and the radiation seems to be frozen for a observer
at the infinite distance.
The density $\rho_\internal(r)$ has the maximum value
$\rho_\maxi = (125 \cdot 3^{3/5})/(4 \pi \cdot 2^{2/5}) \: m_\pl^4/g_*
\simeq 14.57 \times m_\pl^4/g_*$
on the radius $r_{\rhopeak} = 6^{-1/5} r_\BH \simeq 0.6988 \times r_\BH$.
On the same radius, $F(r)$ has the minimum value
$% F(r_\rhopeak) =
F_\mini =
g_* (1-q^2)^2/(200 \cdot 2^{3/10} \cdot 9^{4/5} \sqrt{\pi} m_\pl^2 r_\BH^2)
\simeq 9.515 \times 10^{-4} g_* (1-q^2)^2 /(m_\pl^2 r_\BH^2)$.

On the transitional region
($r_\BH - l_\pl \lnear r \lnear r_\BH + l_\pl$),
the differential equation \EQ{rhoEQ} is approximated to
\begin{eqnarray}
 &&
	\:+\:	r_\BH^3 \rho'^2
		\left\{	\textstyle
		 	\rho
			-9 \frac{Q^2}{8\pi} \frac{1}{r_\BH^4} 
		\right\}
	\:-\:	2 r_\BH^3 \rho  \rho''
		\left\{ \textstyle
			\rho
			-3 \frac{Q^2}{8\pi} \frac{1}{r_\BH^4} 
		\right\}
	\nonumber\\
 &&
	\:+\:
	  \frac{3 m_\pl^2}{8\pi}
	  \left\{
	   + 3 r_\BH \rho'^2  
	   - 2 r_\BH \rho \rho''
	  \right\} = 0.
	  \label{transEQ}
\end{eqnarray}
The approximated form derived by \EQ{transEQ} is a little complicated:
\begin{eqnarray}
 \rho_\trans(r)
  &=&
 	\frac{3}{8\pi} \frac{m_\pl^2}{r_\BH^2} (1 - q^2)
	+
	\frac{(825)^2}{128 \pi^2}
	\frac{m_\pl^4}{g_*}
	\frac{r - r_\BH}{r_\BH^2} \nonumber\\
  &&	\times
	\left[
	      \left( r - r_\BH \right)
	    - \sqrt{ \left(r - r_\BH\right)^2
	             + \frac{96\pi}{(825)^2} \frac{g_*}{m_\pl^2} (1 - q^2)} \:
	\right].
\end{eqnarray}
The element of the metric $G(r)$ has the maximum value $G_\peak$
at the radius $r_{\Gpeak}$
which is quite slightly greater than $r_\BH$ (see \fig{FG.eps}).
In the case of the near extremal ($q\sim1$),
the peak value is approximately given by 
$G_\peak \simeq 50.6 \times g_*^{-1/2} (1 - q^2)^{-3/2} m_\pl r_\BH $
and the radius becomes $r_{\Gpeak} \simeq_\BH$.

\subsection{Singularity of the Radiation-Ball}

When $q \neq 0$,
the asymptotic solution of the singularity ($r=0$) becomes
% \begin{eqnarray}
%  F_\sing(r)
%   &=&
%   \frac{g_*}{720\sqrt{2\pi}}
%   \frac{1}{m_\pl^2 r_\BH^2} \left(\frac{r_\BH}{r}\right),\\
%  G_\sing(r)
%   &=& \frac{g_*}{72 m_\pl^2 r_\BH^2} \left(\frac{r}{r_\BH}\right).
% \end{eqnarray}
\begin{eqnarray}
 F_\sing(r)
  &=&
  \frac{g_* (1 - q^2)^2}{12960 \sqrt{2\pi} m_\pl^4 r_\BH^4} \times
  q^2 \left(\frac{r_\BH}{r}\right)^2, \label{Fsing}\\
 G_\sing(r)
  &=&
  + \frac{1}{q^2} \left(\frac{r}{r_\BH}\right)^2. \label{Gsing}
\end{eqnarray}
$G_\sing(r)$ corresponds to the $r\rightarrow 0$ limit of
the RN metric \EQ{RN-G}.
The power of $r$ in $F_\sing(r)$ corresponds with the RN metric \EQ{RN-F}.
The factor of $F_\sing(r)$ becomes much smaller than that of $F_\BH(r)$
because of the strong red-shift in the ball.
The singularity is time-like, repulsive and naked.
Therefore the singularity qualitatively corresponds
with that of the over-extremal RN metric.

\subsection{Entropy of the Radiation in the Ball}

The total entropy of the radiation in the ball is the same as
that in the chargeless radiation-ball \cite{Nagatani:2003rj}
because $\rho_\internal(r)$ and $G_\internal(r)$ do not depend on $q$.
By combining the entropy density of the radiation
$s = \frac{2\pi^2}{45} g_* T^3$
and the energy density
$\rho = \frac{\pi^2}{30} g_* T^4$,
the entropy density is described as a function of the energy density.
The entropy becomes
\begin{eqnarray}
 S_\internal
  &\simeq&
  \int_{0}^{r_\BH} 4 \pi r^2 \sqrt{G_\internal(r)} \;
  s\left(\rho_\internal(r)\right) \nonumber\\
 &=&
  \frac{(8 \pi)^{3/4}}{\sqrt{5}} m_\pl^2 r_\BH^2
 \;\simeq\; 5.0199 \times \frac{r_\BH^2}{l_\pl^2}.
 \label{entropy-result.eq}
\end{eqnarray}
The entropy \EQ{entropy-result.eq}
is proportional to the surface-area of the ball
and
is a little greater than the Bekenstein entropy \cite{Bekenstein:1973ur}:
\begin{eqnarray}
 S_{\rm Bekenstein}
  &=& \frac{1}{4} \frac{\rm (Horizon\ Area)}{l_\pl^2}
  \;=\; \pi \times \frac{r_\BH^2}{l_\pl^2,}.
\end{eqnarray}
The ratio becomes $S_\internal/S_{\rm Bekenstein} \simeq 1.5978$.
Therefore the origin of the black hole entropy
is regarded as the entropy of the radiation in the ball.
This picture is also effective on the extremal-charged black hole.

In this derivation we have assumed the relativistic thermodynamics
of the radiation.
We expect that treatments of the field dynamics,
e.g. the mode-expansion and the state-counting of the field in the ball,
reproduces the Bekenstein entropy.

\section{Conclusion and Discussion}\label{discussion.sec}

The structure of the radiation-ball
which is identified as the Reissner-Nordstr\"om (RN) black hole
is derived by investigating
the backreaction of the the Hawking radiation into space-time.
There arises no horizon.
The both the outer horizon $r=r_\BH$ 
and the inner horizon $r= q^2 r_\BH$ of the RN black hole
disappear by the backreaction.
The radius of the radiation-ball corresponds to
that of the outer horizon.
The radius of the inner horizon becomes meaningless.
The distribution of the radiation and the metric element $G_\internal(r)$
in the ball do not depend on the charge $q$,
then the area-law of the entropy in the radiation-ball
is not affected by the extension of the charge.
The main effect of the charge appears in the red-shift factor of the ball,
namely, $\sqrt{F(r)_\internal}$ is proportional to $(1-q^2)$.
The Hawking radiation whose temperature is proportion to $(1-q^2)$
is explained as a leak-out of the radiation from the ball with the red-shift.

When we consider the extremal limit $q \rightarrow 1$,
the metric element $F(r)$ goes to zero, i.e.,
the infinite red-shift arises in the ball.
However the structure of the radiation-ball,
including $\rho(r)$ and $G(r)$, is conserved.
Therefore the radiation ball corresponding to
the extremal-charged black hole
seems to be fully frozen with keeping its structure
for the observer at the infinite distance.

In this paper we have assumed that
the only singularity carries the charge and the radiation is neutral.
One may consider extending the ball
such as the radiation carries the charge.
We expect that the extension does not change
the structure of the radiation-ball
because
the energy density of the electromagnetic field is
much smaller than that of the radiation in the ball
except for $r \lnear l_\pl$.
When the radiation carries the hyper charge,
we expect that the ball produces the net baryon number
by the sphaleron process or the GUT interaction
with the baryon-number chemical potential
\cite{CKN, Nagatani:1998gv, Nagatani:2001nz}.
The consideration of the ball with the charged-radiation
is quite interesting
because the phenomenon of the spontaneous charge-transportation
into the black hole is known \cite{Nagatani:2003pr}.

%%%% Acknowledgments %%%%
\begin{flushleft}
 {\Large\bf ACKNOWLEDGMENTS}
\end{flushleft}

 I would like to thank
 Ofer~Aharony, Micha~Berkooz,
 Nadav~Drukker, Bartomeu~Fiol, Hikaru~Kawai, Barak~Kol, Joan~Simon and
 Leonard~Susskind
 for useful discussions.
 I am particularly grateful 
 to Peter Fischer for pointing out a confusing point in the argument.
 I would also like to thank the ITP at Stanford university
 and the organizers of the Stanford-Weizmann workshop
 for their hospitality at the early stage of the project.
 I am grateful to Kei~Shigetomi
 for helpful advice and also for careful reading of the manuscript.
 The work has been supported by
 the Koshland Postdoctoral Fellowship of the Weizmann Institute of Science.

%%%% References %%%%
%\bibliography{references}
%\bibliographystyle{unsrt}

\end{document}